\documentclass[aps,nofootinbib,twocolumn]{revtex4}
\usepackage{amsfonts,amssymb,amsmath}            
\usepackage[]{graphics,graphicx,epsfig}            
\usepackage{amsthm}
\def\identity{\leavevmode\hbox{\small1\kern-3.8pt\normalsize1}}

\newcommand{\ket}[1]{\left | #1 \right\rangle}
\newcommand{\bra}[1]{\left \langle #1 \right |}
\newcommand{\half}{\mbox{$\textstyle \frac{1}{2}$}}

\newcommand{\Tr}{\text{Tr}}

\newcommand{\proj}[1]{\ket{#1}\bra{#1}}
\renewcommand{\epsilon}{\varepsilon}
\begin{document}

\title{Optimal Detection of Entanglement in GHZ States}
\date{\today}

\author{Alastair \surname{Kay}}
\affiliation{Centre for Quantum Computation,
             DAMTP,
             Centre for Mathematical Sciences,
             University of Cambridge,
             Wilberforce Road,
             Cambridge CB3 0WA, UK}
\affiliation{Centre for Quantum Technologies, National University of Singapore, 
			3 Science Drive 2, Singapore 117543}
\begin{abstract}
We present a broad class of states which are diagonal in the basis of $N$-qubit GHZ states such that non-positivity under the partial transpose operation is necessary and sufficient for the presence of entanglement. This class includes many naturally arising instances such as dephased or depolarised GHZ states. Furthermore, our proof directly leads to an entanglement witness which saturates this bound. The witness is applied to thermal GHZ states to prove that the entanglement can be extremely robust to system imperfections.
\end{abstract}

\maketitle

{\em Introduction:} Multipartite entanglement is still a phenomenon that is poorly understood and categorised. For some types of entangled state, such as $N$-qubit GHZ states, it requires very little noise (loss of a single qubit) to entirely destroy the entanglement, whereas others, such as the multipartite states based on error correcting codes, seem much more robust. If we are ever to use entanglement as a resource in information processing protocols, it is essential to understand when entanglement is present in a system and how to detect it. For instance, the ability of entanglement to persist at high temperatures could vastly reduce experimental requirements, and has a direct bearing on the possibility of constructing quantum memories \cite{fernando_prl}.

Attempts to characterise and detect multipartite entanglement are certainly not new. See, for instance, \cite{krauss,dur:99,experiment_witness,otfried,Kay:2006b,Kay:2007,Kay:2008,otfried_rev,otfried_new} and, in particular, the wide-ranging review of \cite{otfried_rev} and the further references contained therein. While the majority of these strategies are capable of detecting some entanglement, the majority, with some notable exceptions including \cite{krauss,dur:99,Kay:2006b,Kay:2007}, are unable to convey how well these characterisations perform. They might be capable of detecting some entanglement, but how much is missed?
 
It seems a reasonable starting point for these studies is to consider classes of states, such as those that are diagonal in the GHZ basis, which arise frequently in quantum information, and are hence likely to be of most interest. Some partial categorisations are already known. For instance, thermal GHZ states can be distilled up to a finite temperature and this temperature is tight i.e.~above that temperature, entanglement can't be distilled \cite{Kay:2006b,Kay:2007}. There is still entanglement present in these models above the distillation threshold \cite{Kay:2008}, which is consequently bound entanglement. D\"ur and Cirac \cite{dur:99} also considered a subset of GHZ-diagonal states and showed that a necessary and sufficient condition for distillation of this subset is that the state should be non-positive with respect to the partial transpose (NPT) operation across every possible bipartition, whereas a necessary and sufficient condition for full separability is that the state should be positive with respect to the partial transpose (PPT) across all possible bipartitions. It does not seem altogether surprising that these two conditions do not, in general, coincide, and hence it is interesting to understand, in as wide a context as possible, how persistent entanglement can be, and how to detect it.

In this paper, we give a broad class of GHZ-diagonal states for which the PPT condition is necessary and sufficient for the state to be fully separable. This class includes the special cases of the thermal state studied in \cite{Kay:2006b,Kay:2007,Kay:2008} and the GHZ-diagonal states in \cite{dur:99}. Our formalism instantly yields an entanglement witness for optimally detecting that entanglement, as well as providing the foundation for a future, more wide-ranging, study of identical concepts in thermal graph states \cite{to_appear}. We describe the special cases of the thermal GHZ state (which is also a dephased GHZ state) and a depolarised GHZ state.

{\em GHZ-diagonal states:} Consider a system of $N$ qubits, and define the stabilizers
$$
K_n=\left\{\begin{array}{cc}
X_1\prod_{m=2}^NZ_m	& n=1	\\
Z_1X_n	& n\geq 2
\end{array}\right.
$$
These terms mutually commute. $X_n$ is the Pauli $X$ matrix applied to qubit $n$, and $Z_x$ denotes the application of $Z$ rotations to all qubits for which the $N$-bit string $x$ is 1, i.e.~by denoting the $n^{th}$ bit of $x\in\{0,1\}^N$ as $x_n$,
$$
Z_x=\prod_{n=1}^NZ_n^{x_n}.
$$
$K_x$ is similarly defined in terms of the $K_n$.
The $+1$ eigenstate of each of these stabilizers is
$$
\ket{\psi}=\ket{0}\ket{+}^{\otimes (N-1)}+\ket{1}\ket{-}^{\otimes (N-1)},
$$
and $\ket{\psi_x}=Z_x\ket{\psi}$ for $x\in\{0,1\}^N$ is an eigenstate of all products of stabilizers $K_y$ with eigenvalue $(-1)^{x\cdot y}$. There is a straightforward local equivalence to any other GHZ basis that one might choose, which does not affect the entanglement structure that we are investigating. However, this particular formulation will allow an immediate translation of many of our results to the more general case of graph states. Any state
\begin{equation}
\rho=\frac{1}{2^N}\sum_{y\in\{0,1\}^N}s_yK_y \label{eqn:gen_form}
\end{equation}
is diagonal in this basis, and it is this class of states which we consider. In order for $\rho$ to be a valid state, we require $s_0=1$ and $\min_{x\in\{0,1\}^N}\sum_ys_y(-1)^{x\cdot y}\geq 0$ (i.e.~the eigenvector with minimum eigenvalue is non-negative).

We are now interested in evaluating the partial transpose criterion on this state in a first step to determine when there is entanglement present in the state. Starting from Eqn.~(\ref{eqn:gen_form}), we introduce a bipartition $z\in\{0,1\}^N$ i.e.~all the vertices with $z_n=0$ are on one side of the partition and those with $z_n=1$ are on the other side, and take the partial transpose on the $z_n=1$ side. Without loss of generality, qubit 1 can be placed on the 0 side of the bipartition and hence $z$ is restricted to being just over the $N-1$ other qubits. Recall that under the partial transpose, the Pauli operators alter by $Z_n\mapsto Z_n$, $X_n\mapsto X_n$ but $Y_n\mapsto(-1)^{z_{n-1}}Y_n$. Thus,
$$
\rho^{PT}=\frac{1}{2^N}\sum_{y\in\{0,1\}^N}s_yK_y(-1)^{y_1\sum_{n=2}^Ny_nz_{n-1}}.
$$
Observe that products of stabilizers remain as products of stabilizers and, as a result, the eigenvectors of $\rho^{PT}$ are just $\ket{\psi_x}$, with eigenvalues $f_{x,z}(\vec s)/2^N$:
$$
f_{x,z}(\vec s)=\sum_{y\in\{0,1\}^N}(-1)^{x\cdot y}s_y(-1)^{y_1\sum_{n=2}^Ny_nz_{n-1}}.
$$
If there exists a choice of $x$ and $z$ such that $f_{x,z}(\vec s)<0$, the state is certainly entangled due to being NPT across the bipartition $z$.

{\em Entanglement Witnesses Saturating PPT:} Using this formalism, it's straightforward to find an entanglement witness that will saturate the PPT threshold for any state which is diagonal in the GHZ state basis. To do this, we measure the observables
\begin{eqnarray}
W_{x,z}&=&\sum_{y\in\{0,1\}^N}(-1)^{x\cdot y}(-1)^{y_1\sum_{n=2}^Ny_nz_{n-1}}K_y.	\nonumber\\
W_{x_1x,z}&\equiv&\!\!\!\!\!\sum_{a\in\{0,1\}}\!\!\!\!\proj{\psi_{ax}}\!+\!(-1)^{x_1+a}\proj{\psi_{a(x\oplus z)}}	\nonumber
\end{eqnarray}
For any arbitrary density matrix $\rho$ with GHZ stabilizer expectation values ${\vec s}$,
$$
\Tr(W_{x,z}\rho)=\!\!\!\!\!\sum_{y\in\{0,1\}^N}\!\!\!(-1)^{x\cdot y}(-1)^{y_1\sum_{n=2}^Ny_nz_{n-1}}s_y=f_{x,z}(\vec s).
$$
Hence, for GHZ-diagonal states, this gives the eigenvalues of the partial transpose of the state about bipartition $z$, and finding $\Tr(W_{x,z}\rho)<0$ for any $x$ or $z$ proves it's entangled. This is a genuine entanglement witness in that, for any state $\rho=\sum_{x,y}\mu_{x,y}\ket{\psi_x}\bra{\psi_y}$, which may not be diagonal in the graph state basis, finding one of the observables to be negative witnesses the fact that it's entangled. To prove this, note that any $\rho$ can be converted, via local probabilistic operations, into a graph diagonal state $\rho_d=\sum_x\mu_{x,x}\proj{\psi_x}$ with the same diagonal elements \cite{aschauer:04}, and hence the same values of ${\vec s}$. So, if $\rho$ is fully separable, it will have the same value of $\Tr(W\rho)$ as $\rho_d$, which we know will be positive since the local conversion to a diagonal state cannot introduce entanglement.

{\em Separability:} We are now in a position where we can determine whether a GHZ-diagonal state is NPT with respect to some bipartition, and have an observable that can witness the entanglement. We now move to studying the converse, when the state is certainly not entangled. Again, the stabilizer formalism is immensely helpful. We will say that $K_x$ and $K_y$ have a {\em compatible basis} if at every site $n$ when $K_x$ is a Pauli matrix $\sigma$, then $K_y$ is either $\sigma$ or $\identity$ at that site and vice versa. We are interested in such cases because each product of stabilizers $K_x$ is just a tensor product of Pauli operators, and hence its eigenvectors are product states. Two terms $K_x$ and $K_y$ have a simultaneous product state decomposition if they have a compatible basis. So, in order to give a fully separable decomposition of $\rho$, we can simply group together all terms that have a compatible basis, and find the smallest eigenvalue. This grouping of terms has to have some component of the $\identity$ added such that the minimum eigenvalue is 0. If we do this, then that grouping of terms is a separable state, with a decomposition specified by the common product basis. We are finally left with a condition that the excess weight of $\identity$ terms should be positive. In the case of a GHZ-diagonal state, the terms $K_y$ for $y\in\{0,1\}^N$ with $y_1=1$ do not have any compatible terms, whereas all terms $K_y$ with $y_1=0$ are mutually compatible. We hence change notation slightly to $K_{1y}$ and $K_{0y}$ respectively for $y\in\{0,1\}^{N-1}$. The decomposition therefore takes the form
\begin{eqnarray}
\rho&=&\sum_{y\in\{0,1\}^{N-1}}|s_{1y}|(\identity+\text{sgn}(s_{1y})K_{1y})	\nonumber\\
&&+\left(\sum_{y}s_{0y}K_{0y}-\identity\min_{x\in\{0,1\}^{N-1}}\sum_{y}s_{0y}(-1)^{x\cdot y}\right)	\nonumber\\
&&+\identity\left(\min_{x\in\{0,1\}^{N-1}}\sum_{y}s_{0y}(-1)^{x\cdot y}-\sum_{y}|s_{1y}|\right).	\nonumber
\end{eqnarray}
Thus, provided
$$
\left(\min_x\sum_{y\in\{0,1\}^{N-1}}s_{0y}(-1)^{x\cdot y}-\sum_{y\in\{0,1\}^{N-1}}|s_{1y}|\right)\geq 0,
$$
we have a separable decomposition of $\rho$. Compare this to $f_{x_0{\tilde x},{\tilde x}\oplus {\tilde z}}(\vec s)$ where $(-1)^{x_0}=-\text{sgn}(s_{100\ldots 0})$, 
$$
\sum_{y\in\{0,1\}^{N-1}}s_{0y}(-1)^{{\tilde x}\cdot y}-\text{sgn}(s_{100\ldots 0})\sum_{y\in\{0,1\}^{N-1}}s_{1y}(-1)^{{\tilde z}\cdot y}.
$$
The two are equal if ${\tilde x}$ corresponds to the minimal choice in the separable state decomposition, and there exists a ${\tilde z}\in\{0,1\}^{N-1}$ such that
\begin{equation}
\text{sgn}(s_{100\ldots 0})s_{1y}(-1)^{{\tilde z}\cdot y}\geq 0\qquad \forall y\in\{0,1\}^{N-1}.	\label{eqn:sufficient}
\end{equation}
(If $s_{100\ldots 0}=0$, then $x_0$ remains a free parameter.) If this simple condition is satisfied, then PPT exactly detects the transition between the existence of bipartite entanglement and full separability of the state which, in turn, makes our entanglement witnesses optimal.

Eqn.~(\ref{eqn:sufficient}) gives a sufficient condition for the coincidence of thresholds for full separability and PPT. If not satisfied, is there really a separation between the PPT threshold and the best known separable state? For a 3-qubit GHZ-diagonal state, Eqn.~(\ref{eqn:sufficient}) is fulfilled provided
$$
\prod_{y\in\{0,1\}^2}s_{1y}\geq 0.
$$
We conclude that roughly half of the parameter space is covered by Eqn.~(\ref{eqn:sufficient}) in this case. One example outside this regime is the state
$$
\rho=\frac{1}{8(1+\alpha)}\left(\prod_{n=1}^3(\identity+K_n)-2K_1K_3+\alpha\identity\right).
$$
Provided $\alpha\geq2$, $\rho$ is a valid state, but it is also PPT with respect to all possible bipartitions. Our previous construction of a separable state is valid for $\alpha\geq 4$. This can be improved to $\alpha\geq 2\sqrt{2}$ by rewriting the sum $K_1+K_1K_2-K_1K_3+K_1K_2K_3$ as
$$
\half\sum_{n=0}^1(X+(-1)^nY)_1(Z+(-1)^nY)_2(Z-(-1)^nY)_3.
$$
Upon implementing the semi-definite programming techniques of \cite{doherty}, we witnessed entanglement in the region $\alpha\leq 2.828$. Hence, the separable decomposition is not universally optimal, but neither is the PPT condition.

\begin{figure}
\begin{center}
\includegraphics[width=0.45\textwidth]{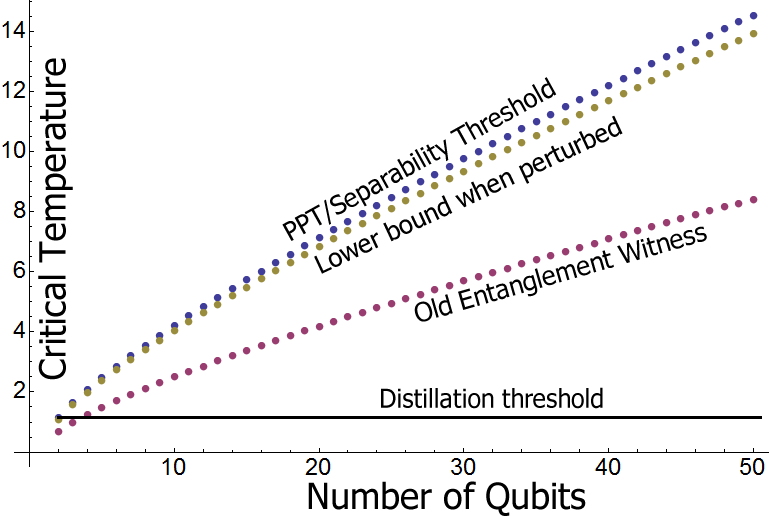}
\vspace{-0.5cm}
\end{center}
\caption{Comparison of the PPT critical temperature for the thermal GHZ state, a lower bound for the model when perturbed by a uniform magnetic field of strength $\delta/\Delta=0.3$, and performance comparison of a previous entanglement witness \cite{Kay:2008}. Choice of $\Delta=k_B=\hbar=1$ ensures unitless quantities.} \label{fig:ghz}
\vspace{-0.5cm}
\end{figure}

{\em Thermal States and Perturbations:} Many noise models satisfy Eqn.~(\ref{eqn:sufficient}), including the sub-class considered in \cite{dur:99} (all $s_{1y}$ equal and positive). We will now discuss two special cases. The first is local dephasing noise on each qubit with probability $p$, and the second will be depolarising noise. Dephasing noise also corresponds to the thermal state of the Hamiltonian
$$
H=-\half\sum_{n=1}^N\Delta_nK_n,
$$
which has $s_y=\prod_{n=1}^N\tanh(\beta\Delta_n/2)^{y_n}$ where $w_y$ is the Hamming weight of $y$, $\beta$ is the inverse temperature, and $\Delta_n$ are energy terms and $p=(1+e^{\beta\Delta})^{-1}$. Since $s_y>0,$ ${\tilde z}=00\ldots 0$. One can also check that ${\tilde x}$ is $11\ldots 1$, and consequently derive a simple threshold condition,
$$
\tanh(\beta\Delta_1/2)=e^{-\beta\sum_{n=2}^N\Delta_n}.
$$
Even though distillable entanglement only persists to a finite temperature \cite{Kay:2006b,Kay:2007}, bound entanglement, which is all bipartite and detected by the partial transpose, persists to a temperature that increases with $N$ (see Fig.~\ref{fig:ghz}). Previous entanglement witnesses, even those specifically designed to detect GHZ state entanglement \cite{Kay:2008}, were often far from optimal. As observed numerically for other graph states in \cite{leandro}, the entanglement is very robust to perturbations in the $\Delta_n$. We can also add a local magnetic field term
$$
H=-\half\sum_{n=1}^N\Delta_nK_n-\half\sum_{n=1}^N\delta_nZ_n.
$$
Since the terms $K_n'=\Delta_nK_n+\delta_nZ_n$ mutually commute, $[K_n',K_m']=0$, and $K_n'^2=(\Delta_n^2+\delta_n^2)\identity$,
$$
\rho=\frac{1}{2^N}\prod_{n=1}^N\left(\identity+\tanh(\half\beta\sqrt{\Delta_n^2+\delta_n^2})K_n'/\sqrt{\Delta^2+\delta^2}\right).
$$
For simplicity of notation, take all the $\Delta_n$ equal, and all the $\delta_n$ equal, although none of the following depends on it. We also set $s=\tanh(\beta\sqrt{\Delta^2+\delta^2}/2)$. The thermal state can be expanded as
$$
2^N\rho=\!\!\sum_{x\in\{0,1\}^N}\sum_{y\in\{0,1\}^N}\!\!\left(\frac{s}{\sqrt{\Delta^2+\delta^2}}\right)^{w_x+w_y}\delta^{w_x}\Delta^{w_y}Z_xK_y
$$
although the summation over $y$ is restricted to cases where $y_n=0$ if $x_n=1$. To prove the presence of entanglement, we can use the entanglement witness $W_{{\tilde x},{\tilde x}\oplus{\tilde z}}$ from the unperturbed case,
$$
\Tr(W_{{\tilde x},{\tilde x}\oplus{\tilde z}}\rho)=\!\!\!\! \sum_{y\in\{0,1\}^N}\!\!\left(\frac{-s\Delta}{\sqrt{\Delta^2+\delta^2}}\right)^{w_y}\!\!(-1)^{y_1\sum_{n=2}^Ny_nz_{n-1}},
$$
since $\Tr(Z_xK_yK_z)=\delta_{x,00\ldots 0}\delta_{y,z}$. The critical inverse temperature $\beta_\delta$ at which the expectation value of this state is zero is hence related to the unperturbed $\beta_0$ by
$$
\frac{\Delta}{\sqrt{\Delta^2+\delta^2}}\tanh(\half\beta_\delta\sqrt{\Delta^2+\delta^2})=\tanh(\half\beta_0\Delta).
$$
Furthermore, $\beta_\delta$ is an upper bound on the true critical $\beta$, i.e.~a lower bound on the critical temperature. Fig.~\ref{fig:ghz} indicates just how robust this entanglement is.

\begin{figure}
\begin{center}
\includegraphics[width=0.45\textwidth]{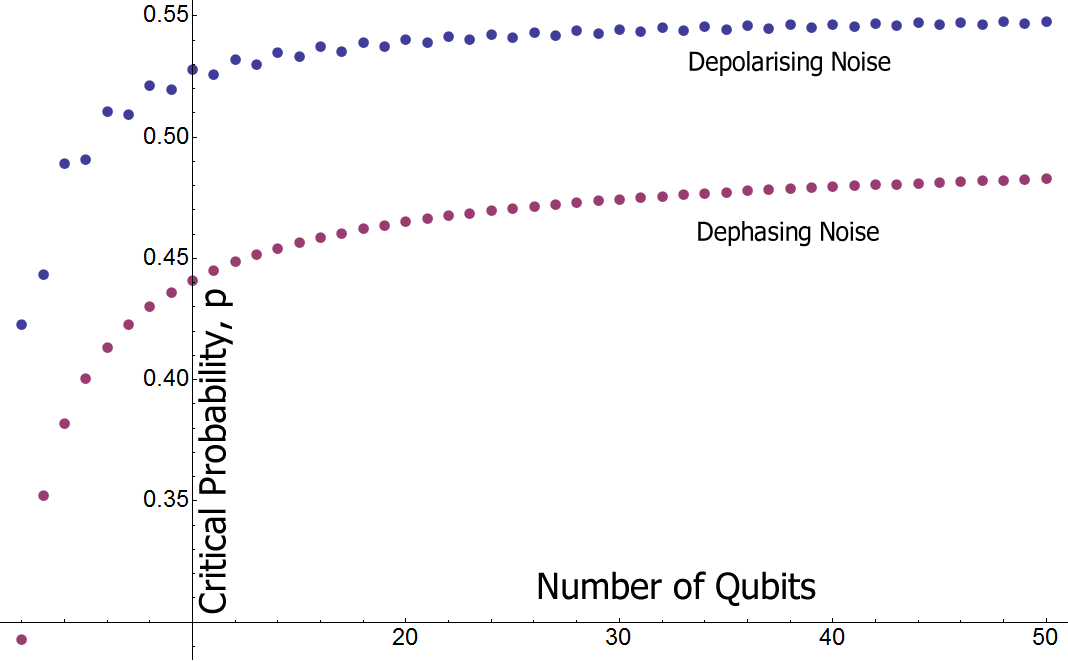}
\vspace{-0.5cm}
\end{center}
\caption{The increasing tolerance of a GHZ state to dephasing and depolarising noise with system size. Note that the two thresholds should not be directly compared since, in the case of depolarising noise, one could argue that an error only happens with a probability $3p/4$, not $p$.}
\label{fig:compare}
\vspace{-0.5cm}
\end{figure}

{\em Depolarising noise:} GHZ states (in our chosen basis) are quite robust against dephasing noise, increasingly so with the number of qubits involved. They are also very tolerant of different noise rates ($\Delta_n$) on different qubits. Another experimentally relevant scenario is when the GHZ state is depolarised independently on each qubit. This is described by a map (on all qubits $n$) of
$$
\mathcal{E}_n(\rho)=(1-p)\rho+\frac{p}{4}\left(\rho+X_n\rho X_n+Y_n\rho Y_n+Z_n\rho Z_n\right).
$$
With a little effort, one can prove that $s_{1y}=(1-p)^N$ and $s_{0y}=(1-p)^{2\lceil w_y/2\rceil}$. Since all $s_{1y}$ are equal, \cite{dur:99} applies for both distillability and separability properties. In the regime of $0\leq p\leq 1$, $s_y\geq 0$ and hence ${\tilde z}=00\ldots 0$. To find ${\tilde x}$, we note that
$$
\min_{x\in\{0,1\}^{N-1}}\sum_{y\in\{0,1\}^{N-1}}(-1)^{x\cdot y}s_{0y}
$$
depends only on the weight of $x$. With this observation in place, one can prove that
$$
\sum_y(-1)^{x\cdot y}s_{0y}=\half(2-p)^{w_x}p^{N-w_x}+\half(2-p)^{N-w_x}p^{w_x},
$$
with a minimum of $(2-p)^{\lfloor N/2\rfloor}p^{\lfloor N/2\rfloor}$ occurring at $w_{\tilde x}=\lfloor N/2\rfloor$. Hence, the PPT threshold and full separability boundary occurs at
$$
(2-p)^{\lfloor N/2\rfloor}p^{\lfloor N/2\rfloor}=2^{N-1}(1-p)^N.
$$
The critical $p$ still increases with $N$, in a similar fashion to the dephasing noise, Fig.~\ref{fig:compare}.

{\em Conclusions:} We have given a sufficient condition, which naturally encompasses a vast range of GHZ-diagonal states, including those that are experimentally relevant, such that the existence of an NPT bipartition is necessary and sufficient for the state to be entangled. We have also shown that there are examples not covered by this condition such that the statement is not true. We have described entanglement witnesses that detect the existence of an NPT bipartition. These are vastly stronger than previous witnesses, and simply correspond to measuring the overlap with four different states. This witness shows that the entanglement in thermal states of GHZ graphs is extremely robust to some classes of perturbation. Future work could focus on witnessing entanglement in those cases where entanglement persists outside the PPT regime, beyond the current reliance on numerical techniques. Criteria developed in \cite{otfried_new} can already prove the existence of entanglement beyond the PPT threshold.

We have been careful to express much of this paper in very general terms using the stabilizer formalism. As such, much of the work extends to graph-diagonal states, which are also defined by stabilizers, and which cover many of the interesting states in quantum information, such as cluster states and error correcting codes. We will study this case in more detail in \cite{to_appear}.

This work was supported by the National Research Foundation \& Ministry of Education, Singapore, and Clare College, Cambridge. O. G\"uhne is thanked for feedback on a previous version of the paper.

\end{document}